\documentclass[aps,prd,reprint,balancelastpage,nofootinbib,amsmath,amssymb,superscriptaddress,longbibliography]{revtex4-1}
\usepackage[
colorlinks=true,        
citecolor=blue,         
linkcolor=blue,         
urlcolor=blue           
]{hyperref}             
 \usepackage{graphicx}
\usepackage{bm}         

\newcommand{\nc}{\newcommand*} 
\nc{\Eq}[1]{Eq.~\eqref{#1}}     
\nc{\Fig}[1]{Fig.~\ref{#1}}     
\nc{\Table}[1]{Table~\ref{#1}}  
\nc{\Sec}[1]{Sec.~\ref{#1}}     
\nc{\red}[1]{\textcolor{red}{#1}}
\nc{\dt}{\delta}
\nc{\bt}{\mathbf{t}}
\nc{\eg}{\textit{e.g.~}}
\nc{\bn}[1]{\dt\bm{t}_{\text{#1}}}
\nc{\be}{\bm{\epsilon}}
\nc{\BF}{\mathcal{BF}}
\nc{\yr}{\mathrm{yr}}

\def\e{\begin{equation}}
\def\q{\end{equation}}

\begin{document}
	
\title{Infrared Behavior of Induced Gravitational Waves from Isocurvature Perturbations}
\author{Chang Han}
\affiliation{Department of Physics and Synergetic Innovation Center for Quantum Effects and Applications, Hunan Normal University, Changsha, Hunan 410081, China}
\affiliation{Institute of Interdisciplinary Studies, Hunan Normal University, Changsha, Hunan 410081, China}

\author{Zu-Cheng Chen}
\email{zuchengchen@hunnu.edu.cn}
\affiliation{Department of Physics and Synergetic Innovation Center for Quantum Effects and Applications, Hunan Normal University, Changsha, Hunan 410081, China}
\affiliation{Institute of Interdisciplinary Studies, Hunan Normal University, Changsha, Hunan 410081, China}

\author{Hongwei Yu}
\email{hwyu@hunnu.edu.cn}
\affiliation{Department of Physics and Synergetic Innovation Center for Quantum Effects and Applications, Hunan Normal University, Changsha, Hunan 410081, China}
\affiliation{Institute of Interdisciplinary Studies, Hunan Normal University, Changsha, Hunan 410081, China}

\author{Puxun Wu}
\email{pxwu@hunnu.edu.cn}
\affiliation{Department of Physics and Synergetic Innovation Center for Quantum Effects and Applications, Hunan Normal University, Changsha, Hunan 410081, China}
\affiliation{Institute of Interdisciplinary Studies, Hunan Normal University, Changsha, Hunan 410081, China}


\begin{abstract}
Induced gravitational waves provide a powerful probe of primordial perturbations in the early universe through their distinctive spectral properties. We analyze the spectral energy density $\Omega_{\text{GW}}$ of gravitational waves induced by isocurvature scalar perturbations. 
In the infrared regime, we find that the spectral slope $n_{\text{GW}} \equiv \text{d} \ln\Omega_\mathrm{GW}/\text{d}\ln k$ takes the log-dependent form $3-4/ \ln (\tilde{k}_*^2 / 6k^2)$, where $\tilde{k}_*$ represents the effective peak scale of the primordial scalar power spectrum. This characteristic behavior differs markedly from that of adiabatic-induced gravitational waves, establishing a robust observational discriminant between isocurvature and adiabatic primordial perturbation modes.
\end{abstract}

\maketitle
\section{Introduction}

The stochastic gravitational wave background (SGWB) serves as a unique window into early universe physics, particularly through its ability to probe primordial perturbations that are inaccessible through the electromagnetic observations~\cite{WMAP:2003ivt,Planck:2018jri}. Of particular interest are primordial scalar perturbations, which can generate significant gravitational wave (GW) signatures through second-order effects~\cite{Ananda:2006af,Baumann:2007zm}. When sufficiently amplified, these perturbations may produce induced GWs (IGWs) in the frequency range $10^{-9}$--$10^3$ Hz and lead to the formation of primordial black holes (PBHs)~\cite{Hawking:1971ei,Carr:1974nx,Sasaki:2018dmp}, potentially contributing to both dark matter abundance and the GW events detected by LIGO-Virgo-KAGRA~\cite{Bird:2016dcv,Sasaki:2016jop,Chen:2024dxh}.

Primordial fluctuations manifest in two distinct modes: adiabatic and isocurvature~\cite{Kodama:1984ziu,Bucher:1999re}. Adiabatic fluctuations emerge when all components share a common spacetime slicing where their separate energy density fluctuations vanish, leaving only metric perturbations. This behavior typically results from single-field inflation models and has been confirmed by cosmic microwave background (CMB) observations~\cite{WMAP:2003ivt,Planck:2018jri}, with an amplitude of approximately $10^{-9}$ on large scales.

Isocurvature perturbations, in contrast, represent variations in relative number densities between different species while maintaining homogeneous total energy density~\cite{Kodama:1984ziu,Bucher:1999re}. These fluctuations arise naturally in multi-field inflation models~\cite{Chung:2017uzc,Chung:2021lfg}, phase transitions~\cite{Dolgov:1992pu}, and from Poisson noise in early universe structure formation~\cite{Inman:2019wvr,Papanikolaou:2020qtd}. While CMB measurements constrain large-scale isocurvature contributions to less than 1-10\% of total fluctuations~\cite{Planck:2018jri}, their behavior on smaller scales remains largely unconstrained.

The generation of IGWs occurs through the coupling between first-order scalar perturbations and second-order tensor perturbations during the radiation-dominated era~\cite{Mukhanov:2005sc,Malik:2008im}. For adiabatic perturbations, this process is well understood within the standard cosmological model. However, isocurvature perturbations introduce additional complexity through the interaction between multiple fields~\cite{Domenech:2021wkk}. These fluctuations influence the evolution of the energy-momentum tensor as component energy density ratios vary, necessitating careful consideration of the isocurvature slicing where primordial metric fluctuations vanish~\cite{WMAP:2003elm}.

The distinct properties of adiabatic and isocurvature modes raise the prospect of using their observational signatures for discriminating between different early universe scenarios. While CMB spectral distortions provide access to isocurvature modes in the range $1\,{\rm Mpc}^{-1} \lesssim k \lesssim 10^6\,{\rm Mpc}^{-1}$~\cite{Chluba:2013dna,Chluba:2019kpb}, probing smaller scales requires alternative approaches such as PBHs and IGWs. The GW spectrum is particularly promising for this discrimination, as these modes are expected to produce distinct signatures~\cite{Domenech:2021ztg}. Recent pulsar timing array observations have already opened new possibilities for testing these differences~\cite{Chen:2024twp}. For adiabatic perturbations, detailed theoretical work has shown that the infrared behavior of the GW spectrum exhibits a log-dependent slope of $n_{\text{GW}}=3-4/\ln(4k_*^2/3k^2)$~\cite{Yuan:2019wwo}, which approaches $n_{\text{GW}}=3$~\cite{Cai:2019cdl} in the limit $k/k_* \to 0$, where $k_*$ is the characteristic wavenumber corresponding to the peak of the scalar perturbation power spectrum. However, the corresponding spectral behavior for the isocurvature mode remains unexplored.

In this work, we analyze the spectral properties of GWs induced by isocurvature fluctuations, focusing particularly on their infrared behavior. We derive analytical expressions for the spectral slopes and compare them with the adiabatic case, demonstrating how these distinct features can serve as discriminators between fundamental modes of primordial perturbations. The paper is organized as follows. In \Sec{SIGW}, we review the calculation of IGW from isocurvature perturbations and derive the GW energy density spectrum $\Omega_{\text{GW}}$. In \Sec{Slope}, we analyze the infrared behavior of $\Omega_{\text{GW}}$ for the isocurvature mode and compare with the adiabatic mode, focusing on their distinct spectral slopes. Finally, in \Sec{conclusion}, we summarize our findings.

\section{Induced gravitational waves}\label{SIGW}

In this section, we review the formalism for calculating IGWs from isocurvature perturbations. 
We consider perturbations to a FLRW metric in the conformal Newtonian gauge~\cite{Ananda:2006af}
\begin{equation}
    \label{eq:gauge}
    ds^2=a^2\left\{-(1+2\Psi)d\tau^2+\left[(1-2\Psi)\delta_{ij}+\frac{h_{ij}}{2}\right]dx^idx^j\right\},
\end{equation}
where $\tau$ is conformal time, $a(\tau)$ is the scale factor, $\Psi$ represents the scalar perturbation in the gravitational potential, and $h_{ij}$ denotes tensor perturbations satisfying transverse-traceless conditions.

The multi-component energy-momentum tensor is given by
\begin{equation}
    \label{eq:Tmunu}
    T_{\mu\nu}=(\rho_r+\rho_m+p_r)u_\mu u_\nu -p\delta_{\mu\nu},
\end{equation}
where subscripts $r$ and $m$ denote radiation and matter components respectively. In this multi-fluid system, the pressure perturbations can be decomposed into adiabatic and non-adiabatic parts as~\cite{Domenech:2023jve,Mukhanov:2005sc}
\begin{equation}
    \label{eq:deltaPu}
    \delta p=c_s^2\, \delta\!\rho_r+ c_s^2\rho_m  S,
\end{equation}
where $c_s$ is the sound speed and $S$ represents the entropy perturbation fraction, defined as the relative fluctuation in number densities:
\begin{equation}
    \label{eq:defS}
    S\equiv \frac{\delta(\frac{n_r}{n_m})}{\frac{n_r}{n_m}}=\frac{3}{4}\frac{\delta\!\rho_r}{\rho_r}-\frac{\delta\!\rho_m}{\rho_m}.
\end{equation}
In the early universe, radiation and matter are tightly coupled, allowing us to treat $S$ as a frequency-dependent but time-independent quantity.

In Fourier space, the evolution equation for scalar perturbations with comoving wavenumber $k$ takes the form
\begin{equation}\label{eq:eqPhiS}
\begin{aligned}
    \Psi_k''+(1+3c_s^2)\mathcal{H}^2 \Psi_k+2\mathcal{H} '\Psi_k+3(1+c_s^2)\mathcal{H}\Psi_k'&+c_s^2k^2\Psi_k
    \\=&\frac{a^2}{2}c_s^2\rho_m S_k,
\end{aligned}
\end{equation}
where prime denotes derivatives with respect to $\tau$, and $\mathcal{H}=a'/a$ is the conformal Hubble parameter. The solution for the isocurvature mode is given by~\cite{Domenech:2023jve}
\begin{equation}\label{PhiS}
\begin{aligned}
    \Psi_{k }(\tau)\simeq&
    \frac{3S_k}{2\sqrt{2}}\frac{k_{\text{eq}}}{k}\frac{1}{(k\tau)^3}
    \\& \times\left[6+(k\tau)^2-2\sqrt{3}k\tau \sin(c_sk\tau)-6\cos(c_sk\tau)\right]  ,
\end{aligned}
\end{equation}
where $k_{\text{eq}}$ corresponds to the horizon scale at matter-radiation equality.
The statistical properties of isocurvature perturbations are characterized by their power spectrum $\mathcal{P}(k)$, defined through
\begin{equation}
\left \langle S_kS_{k'} \right \rangle=\frac{2\pi^2}{k^3}\mathcal{P}(k)\times (2\pi)^3\delta ^{(3)}(k+k').
\end{equation}

The tensor perturbations $h_{ij}$ evolve according to
\begin{equation}
       h''_{ij}+2\mathcal{H}h'_{ij}-\nabla ^2h_{ij}=-4T^{lm}_{ij}\mathcal{S} _{lm},
\end{equation}
where $T^{lm}{ij}$ is the transverse-traceless projection operator. The source term $\mathcal{S}_{ij}$ contains quadratic combinations of scalar perturbations:
\begin{equation}
\begin{aligned}
    \mathcal{S} _{ij}
         =& 4\Psi \partial_i \partial_j \Psi + 2\partial_i \Psi \partial_j \Psi 
         \\&-3c_s^2\frac{\rho_r+\rho_m}{\rho_r}\frac{1}{\mathcal{H}^2}\partial_i(\mathcal{H}\Psi+\Psi')\partial_j(\mathcal{H}\Psi+\Psi').
\end{aligned}
\end{equation}

The present-day energy density spectrum of IGWs can be expressed as
\begin{equation}
    \Omega_{\text{GW}}(k)=\Omega_r\int^\infty_0 dv \int^{1+v}_{|1-v|} du I(u,v,k) \mathcal{P}(vk)\mathcal{P}(uk),
\end{equation}
where $\Omega_r$ is the present radiation density parameter. During radiation domination, $c_s^2=1/3$, the kernel function takes the form~\cite{Domenech:2023jve}
\begin{equation}\label{eq:Iuvk}
\begin{aligned}
    I(u,v,k)=&\frac{1}{12}\left[\frac{4v^2-(1+v^2-u^2)^2}{4uv}\right]^2\left(\frac{k_{\text{eq}}}{k}\right)^4
           \\&\times\frac{1}{2}\left[I_c^2(u,v)+I_s^2(u,v)\right],
\end{aligned}
\end{equation}
where $I_s(u,v)$ and $I_c(u,v)$ are given by
\begin{equation}
\begin{aligned}
     I_s (u,v)=& \frac{9\pi}{32u^3v^3}\Big [27-18u^2+u^4+6(-3+u^2)v^2+v^4
         \\&-2(-3+u^2)(-3+u^2+2v^2)\Theta\left(1-\frac{u}{\sqrt{3}}\right)
         \\&-2(-3+v^2)(-3+2u^2+v^2)\Theta\left(1-\frac{v}{\sqrt{3}}\right)
         \\&+(-3+u^2+v^2)^2\Theta\left(1-\frac{u}{\sqrt{3}}-\frac{v}{\sqrt{3}}\right)\Big ],
\end{aligned}
\end{equation}
and
\begin{equation}
    \begin{aligned}
        I_c(u,v)=& \frac{9}{32u^3v^3}\Big [-6u^2v^2
         \\&  +2(-3+u^2)(-3+u^2+2v^2)\ln\left|\frac{-3+u^2}{3}\right|
        \\&+2(-3+v^2)(-3+2u^2+v^2)\ln\left|\frac{-3+v^2}{3}\right|
        \\&-\left(-3+u^2+v^2\right)^2
        \\&\quad\times\ln\left|\frac{u^4+(-3+v^2)^2-2u^2(3+v^2)}{9}\right|\Big ].
    \end{aligned}
\end{equation}
Here, $u$ and $v$ are dimensionless variables and $\Theta$ denotes the Heaviside function.

For analytical convenience, we absorb the explicit $k$ dependence in \Eq{eq:Iuvk} by defining the effective power spectrum as
\begin{equation}\label{Ptilde}
    \widetilde{\mathcal{P}}(k)=\left(\frac{k_{\text{eq}}}{k}\right)^2\mathcal{P}(k).
\end{equation}
This allows us to write the GW energy density in the more compact form:
\begin{equation}\label{eq:Omegake0}
    \Omega_{\text{GW}}(k)=\Omega_{r}\int^\infty_0 dv \int^{1+v}_{|1-v|} du \tilde{I}(u,v) \widetilde{\mathcal{P}}(uk)\widetilde{\mathcal{P}}(vk),
\end{equation}
where the $k$-independent kernel function $\tilde{I}(u,v)$ is given by
\begin{equation}
\begin{aligned}
    \tilde{I}(u,v)=&\frac{1}{24}\left[\frac{4v^2-(1+v^2-u^2)^2}{4uv}\right]^2
           \left[I_c^2(u,v)+I_s^2(u,v)\right].
\end{aligned}
\end{equation}

\section{Slope of Induced Gravitational Waves}\label{Slope}

The spectral shape of IGWs provides a distinctive signature for discriminating between primordial fluctuation modes, particularly between isocurvature and adiabatic perturbations. Following the framework developed in Ref.~\cite{Yuan:2019wwo}, we analyze the IGW spectral slope arising from isocurvature modes, with special emphasis on the infrared regime where the distinguishing features become most apparent.

We consider an effective power spectrum $\widetilde{\mathcal{P}}(k)$ that 
exhibits a pronounced peak at $\tilde{k}_*$ and is non-zero only within the interval $k_-<k<k_+$. This effective spectrum relates to the original power spectrum $\mathcal{P}(k)$ through Eq.~\eqref{Ptilde}, where the factor $(k_{\text{eq}}/k)^2$ ensures that $\tilde{k}_*\leq k_*$. 
For physical consistency, we restrict our analysis to $k>k_{\text{eq}}$, where $k_{\text{eq}}$ corresponds to the matter-radiation equality scale. This restriction naturally excludes CMB scales, which lie outside our frequency range of interest, and prevents infrared divergences. The GW energy density for this effective power spectrum takes the form
\begin{equation}\label{ogw2}
\Omega_{\text{GW}}(k)=\Omega_{r}\!\int^\frac{k_+}{k}_\frac{k_-}{k}\! dv\! \!\int^{u_2(v)}_{u_1(v)}\!du\, \tilde{I}(u,v)\widetilde{\mathcal{P}}(vk)\widetilde{\mathcal{P}}(uk),
\end{equation}
where the integration limits for $u$ are $u_1(v) \equiv \max\{v-1,k_-/k\}$ and $u_2(v) \equiv \min\{v+1,k_+/k\}$. 

\subsection{Infrared Region Analysis}
In the infrared limit ($k \ll k_-$), both integration variables satisfy $u,v\gg1$. For a spectrum satisfying $k_-/k < v-1 < v+1 < k_+/k$, the integration range for $u$ spans $[v-1,v+1]$, yielding
\begin{equation}
\Omega_{\text{GW}}(k)=\Omega_{r}\!\int^\frac{k_+}{k}_\frac{k_-}{k}\! dv \!\int^{v+1}_{v-1}\! du\, \tilde{I}(u,v) \widetilde{\mathcal{P}}(uk)\widetilde{\mathcal{P}}(vk).
\end{equation}
The kernel function in this limit takes the form
\begin{equation}
\label{eq:I1}
\begin{aligned}
    \tilde{I}(u,v)\simeq &\frac{27 \left [(u^2-v^2)^2-4v^2\right]^2}{(4uv)^8} \times \Big[2(u^4+2u^2v^2)\ln\frac{u^2}{3}\\
 &+ 2(v^4+2u^2v^2)\ln\frac{v^2}{3} -(u^2+v^2)^2\ln\frac{(u+v)^2}{3} \Big ]^2.
\end{aligned}
\end{equation}
For large $v$, the $u$-integral can be approximated as
\begin{equation}
\int^{v+1}_{v-1} du\, \tilde{I}(u,v) \widetilde{\mathcal{P}}(vk)\widetilde{\mathcal{P}}(uk)    \simeq 2\tilde{I}(v,v){\widetilde{\mathcal{P}}}^2(vk),
\end{equation}
where
\begin{equation}\label{eq:I2}
    \tilde{I}(v,v)\simeq \frac{27\ln^2\frac{v^2}{6}}{64v^4}.
\end{equation}

In the infrared limit ($k \ll \tilde{k}_*$), the dominant contribution to the $v$-integral arises from the region below the peak, allowing us to further approximate the integration range as $k_-/k < v < \tilde{k}_*/k$. By introducing the dimensionless variable $y \equiv vk/\tilde{k}_*-1$, we obtain the range $-1 < y < 1$. This enables a small-$y$ expansion, yielding the scaling behavior
\begin{equation}
    \Omega_{\text{GW}}(k)\propto \left(\frac{\tilde{k}_*}{k}\right)^3\ln^2\frac{{\tilde{k}_*}^2}{6k^2},
\end{equation}
from which we derive the GW spectral slope as
\begin{equation}\label{eq:nGW3} 
n_{\text{GW}}\equiv \frac{\text{d} \ln\Omega_\text{GW}}{\text{d}\ln k}=3-\frac{4}{\ln\frac{{\tilde{k}_*}^2}{6k^2}}.
\end{equation}
While the slope depends on the specific form of the primordial power spectrum through the effective peak scale $\tilde{k}_*$, we find that in the deep infrared limit ($k/\tilde{k}_* \to 0$), the spectral slope universally approaches $n_{\text{GW}} \to 3$, independent of the detailed shape of the power spectrum.

\subsection{Near-peak Region Analysis}

While \Eq{eq:nGW3} describes the infrared behavior, we can obtain more precise results near the peak $\tilde{k}_*$ for narrow spectra where $\Delta \equiv (k_+ - k_-)/\tilde{k}_* \ll 1$. In this regime, $\tilde{k}_* \simeq k_*$, and the integration limits simplify considerably when $(k_+/k-1)< k_-/k < k_+/k < (k_-/k+1)$. Under these conditions, \Eq{ogw2} becomes
\begin{equation}
\Omega_{\text{GW}}(k)=\Omega_{r}\int^\frac{k_+}{k}_\frac{k_-}{k} dv \int^\frac{k_+}{k}_\frac{k_-}{k} du \tilde{I}(u,v) \widetilde{\mathcal{P}}(vk)\widetilde{\mathcal{P}}(uk).
\end{equation}

\begin{figure}[tbp]
\centering
\includegraphics[width=0.48\textwidth]{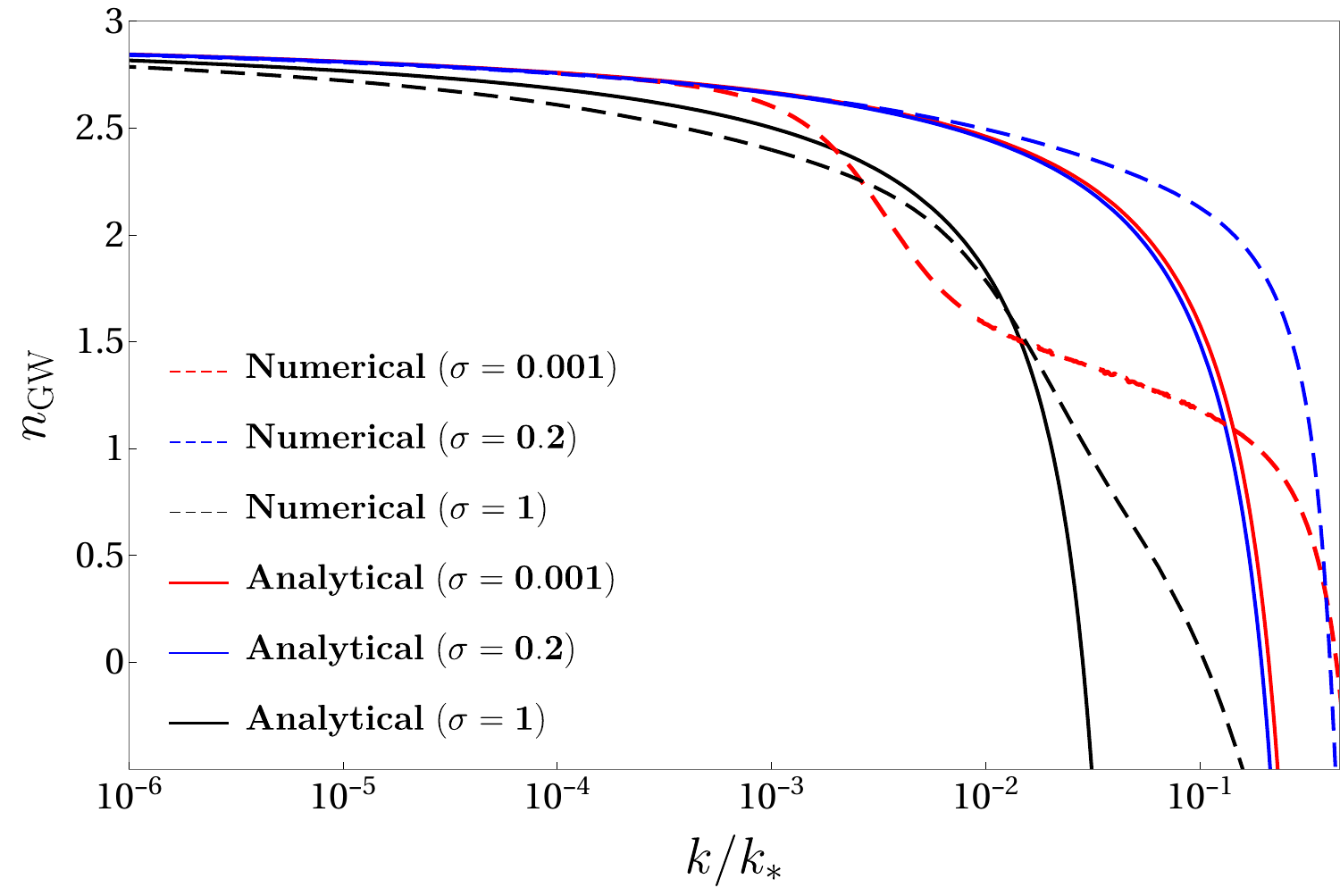}
\caption{\label{fig:lognormal}GW spectral slopes $n_{\text{GW}}$ induced by lognormal power spectra of varying widths $\sigma$. Dashed lines show numerical results for narrow ($\sigma=0.001$, red), intermediate ($\sigma=0.2$, blue), and wide ($\sigma=1$, black) spectra. Solid lines represent the corresponding analytical predictions from Eq.~\eqref{eq:nGW3}.}
\end{figure}

For this near-peak analysis, we need a refined approximation of the kernel function that retains more terms than Eq.~\eqref{eq:I1}:
\begin{equation}\label{eq:I3}
\begin{aligned}
    \tilde{I}(u,v)\simeq &\frac{27
\left [(u^2-v^2)^2-4v^2\right]^2
}{(4uv)^8}\\
 &\times \Big\{\pi^2(u^4+v^4+6u^2v^2-18v^2-18u^2)^2\\
&+\Big[2(u^2-3)(u^2+2v^2-3)\ln\frac{u^2}{3}\\
&+ 2(v^2-3)(v^2+2u^2-3)\ln\frac{v^2}{3}\\
&-6u^2v^2-(u^2+v^2-3)^2\ln\frac{(u+v)^2}{3} \Big]^2 \Big \}.
\end{aligned}
\end{equation}
By introducing the dimensionless variables $x\equiv uk/k_*-1$ and $y\equiv vk/k_*-1$ and performing a perturbative expansion around $x,y \approx 0$ while neglecting higher-order terms, we obtain
\begin{equation}
\begin{aligned}
     &\Omega_{\text{GW}}(k)\propto\frac{k^2}{k_*^{10}}\Big\{16k_*^2(9k^2-2k_*^2)^2\pi^2\\ 
     &~+\Big[6k_*^4-12\left(3k^4-4k^2k_*^2+k_*^4\right)\ln\frac{k_*^2}{3k^2} \\
     &~+\left(3k^2-2k_*^2\right)^2\ln\frac{4k_*^2}{3k^2}\Big]^2\Big\}.
\end{aligned}   
\end{equation}
The corresponding GW spectral slope follows
\begin{equation}\label{eq:nGWimprove}
\begin{aligned}
n_{\text{GW}}=2-8\frac{36\pi^2-\left(3-4\ln\frac{k_*^2}{6k^2}\right)
\left(2\frac{k_*^2}{k^2}+9\ln\frac{k_*^2}{2^{2/3}3k^2}\right)}{\frac{k_*^2}{k^2} \left[16\pi^2+\left (3-4\ln\frac{k_*^2}{6k^2}\right )^2\right]}.
\end{aligned}
\end{equation}
Notably, this result is independent of the specific functional form of the power spectrum.

\subsection{Two Concrete Examples}

To validate our analytical results, we examine two representative power spectrum models commonly used in the literature~\cite{Bartolo:2018rku}.

\begin{figure}[tbp]
\centering
\includegraphics[width=0.48\textwidth]{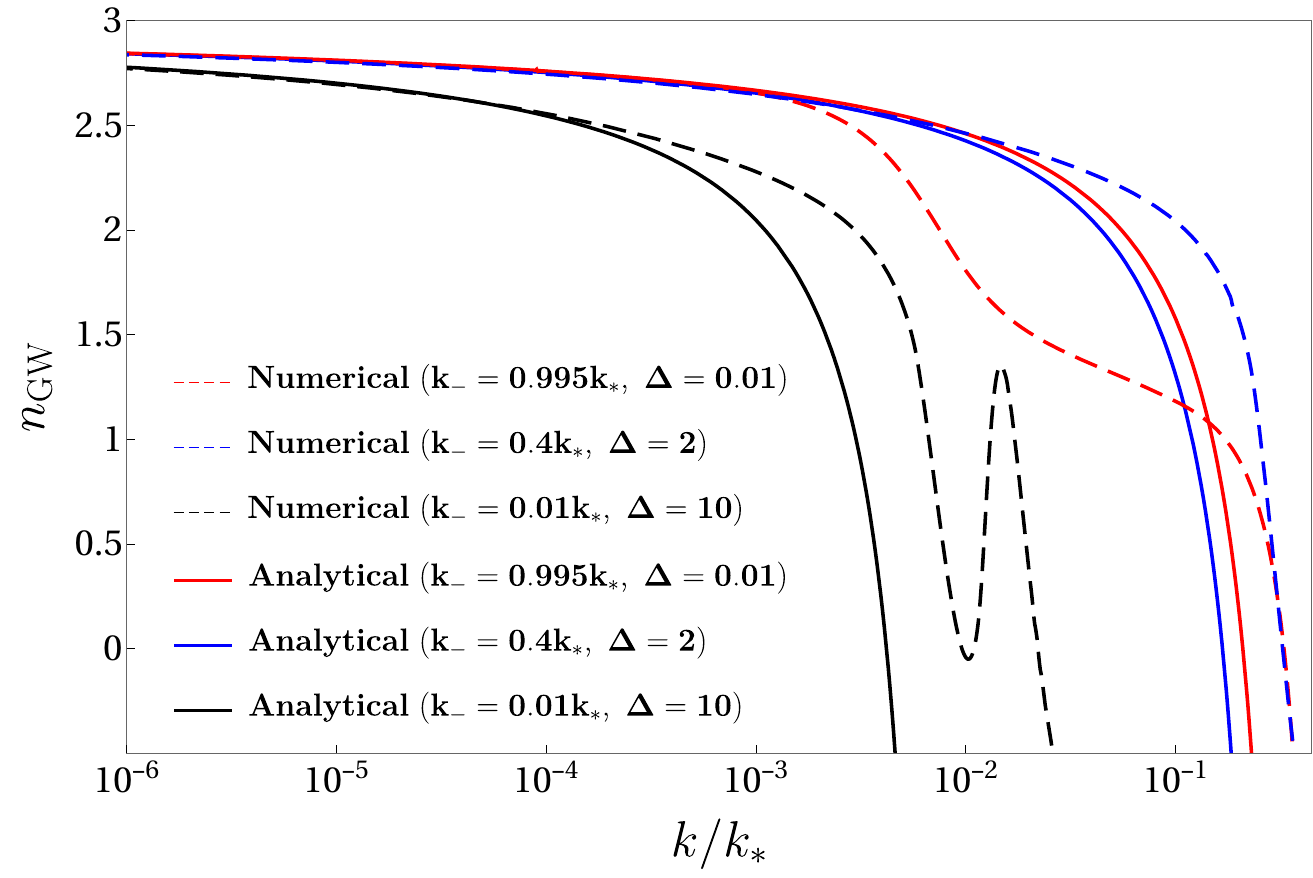}
\caption{\label{fig:broken}GW spectral slopes $n_{\text{GW}}$ induced by broken power spectra with varying spectral widths $\Delta$. Dashed lines show numerical results for narrow ($k_-=0.995k_*$, $\Delta=0.01$, red), intermediate ($k_-=0.4k_*$, $\Delta=2$, blue), and wide ($k_-=0.01k_*$, $\Delta=10$, black) spectra. Solid lines represent the corresponding analytical predictions from Eq.~\eqref{eq:nGW3}.}
\end{figure}

First, we consider a lognormal power spectrum
\begin{equation}\label{eq:lognormal}
    \mathcal{P}(k)=\frac{\mathcal{A}}{\sqrt{2\pi\sigma^2}}e^{-\frac{\ln\left(\frac{k}{k_*}\right)^2}{2\sigma^2}},
\end{equation}
which yields the effective power spectrum
\begin{equation}
    \widetilde{\mathcal{P}}(k)=\left(\frac{k_{\text{eq}}}{k}\right)^2 \frac{\mathcal{A}}{\sqrt{2\pi\sigma^2}}e^{-\frac{\ln\left(\frac{k}{k_*}\right)^2}{2\sigma^2}}.
\end{equation}
Here, $\mathcal{A}$ represents the amplitude and $\sigma$ characterizes the logarithmic width of the peak. By taking the derivative with respect to $k$ to maximize $\widetilde{\mathcal{P}}(k)$, we find
\begin{equation}\label{eq:ke1}
\tilde{k}_*= e^{-2\sigma^2}k_*.
\end{equation}

Second, we analyze a broken power spectrum
\begin{equation}
\label{eq:broken}
     \mathcal{P}(k)=  \mathcal{A}\times \begin{cases}
    \frac{k-k_-}{k_*-k_-}, &\text{ for } k_-<k<k_*, \\
    \frac{k_+-k}{k_+-k_*}, &\text{ for } k_*<k<k_+,\end{cases}
\end{equation}
with corresponding effective spectrum
\begin{equation}
    \widetilde{\mathcal{P}}(k)=   \mathcal{A}\left(\frac{k_{\text{eq}}}{k}\right)^2\times \begin{cases}
    \frac{k-k_-}{k_*-k_-}, &\text{ for } k_-<k<k_*, \\
    \frac{k_+-k}{k_+-k_*}, &\text{ for } k_*<k<k_+,\end{cases}
\end{equation}
leading to
\begin{equation}
\label{eq:ke2}
\tilde{k}_*=\text{min}\{2k_-,k_*\}.
\end{equation}

\begin{figure}[t]
\centering
\includegraphics[width=0.48\textwidth]{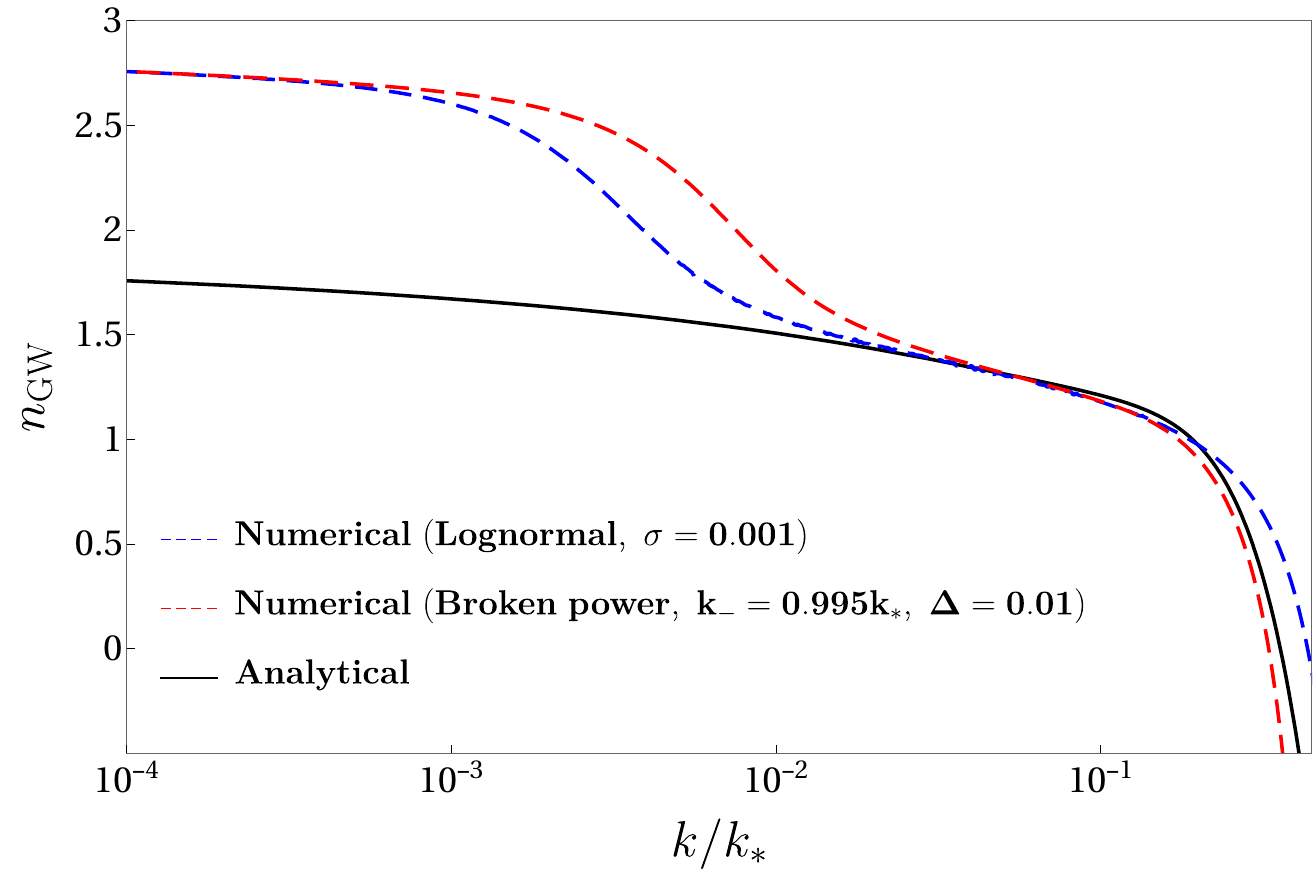}
\caption{\label{fig:improve}GW spectral slopes $n_{\text{GW}}$ induced by narrow primordial scalar power spectra. Dashed lines show numerical results for a lognormal spectrum ($\sigma=0.001$, red) and a broken power spectrum ($k_-=0.995k_*$, $\Delta=0.01$, blue). Solid lines represent the refined analytical predictions from Eq.~\eqref{eq:nGWimprove}, demonstrating enhanced accuracy near the peak frequency.}
\end{figure}

Our numerical analysis, presented in Figs.~\ref{fig:lognormal}-\ref{fig:improve}, validates the analytical framework across multiple parameter regimes. \Fig{fig:lognormal} examines lognormal spectra with varying widths ($\sigma=0.001, 0.2,$ and $1$), while \Fig{fig:broken} studies broken power spectra with different characteristic scales ($\Delta=0.01, 2,$ and $10$). Both figures demonstrate excellent agreement between analytical predictions (solid lines) and numerical calculations (dashed lines) in the infrared region. For narrow spectra ($\sigma=0.001$ for lognormal and $\Delta=0.01$ for broken power), \Fig{fig:improve} confirms the validity of our refined near-peak analysis.

\begin{figure}[t]
\centering
\includegraphics[width=0.48\textwidth]{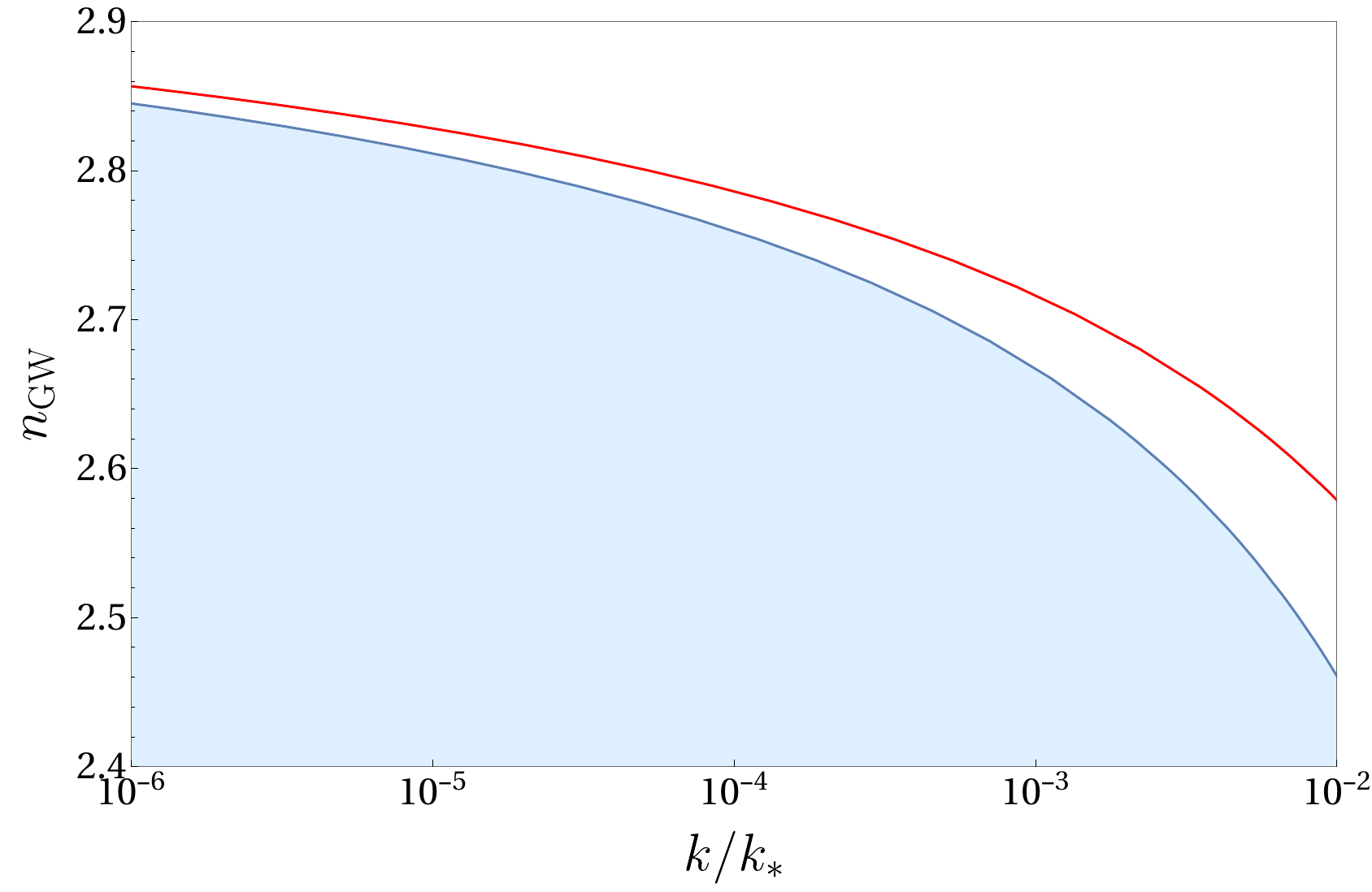}
\caption{\label{fig:compare}Comparison of GW spectral slopes $n_{\text{GW}}$ induced by adiabatic and isocurvature perturbation modes. The red curve shows the adiabatic mode behavior $n_{\text{GW}}=3-4/\ln(4k_*^2/3k^2)$~\cite{Yuan:2019wwo}, while the blue shaded region represents the accessible range of isocurvature slopes from Eq.~\eqref{eq:nGWb} for $0 \leq \alpha \leq 1$. The systematic separation between these spectral behaviors provides a robust discriminant for the underlying perturbation modes.}
\end{figure}

While the infrared behavior of $n_{\text{GW}}$ depends on the power spectrum $\mathcal{P}(k)$ through $\tilde{k}_*$, this dependence can be parameterized using the relation $\tilde{k}_* = \alpha k_*$, where $0<\alpha\lesssim 1$ since $\tilde{k}_* \lesssim k_*$. Under this parameterization, \Eq{eq:nGW3} becomes
\begin{equation}
\label{eq:nGWb}
n_{\text{GW}}= 3-\frac{4}{\ln\frac{{(\alpha k_*)}^2}{6k^2}}.
\end{equation}
This result stands in marked contrast to the adiabatic case, where~\cite{Yuan:2019wwo}
\begin{equation}\label{nGW35}
    n_{\text{GW}}=n_0-\frac{4}{\ln\frac{4k_*^2}{3k^2}}
\end{equation}
with $n_0=3$, which is independent of the underlying power spectrum shape.   For extremely narrow spectra near the peak, it has been found $n_0=2$~\cite{Yuan:2019wwo}. 
The distinction between  the infrared behaviors of GW spectral slopes induced by  adiabatic and isocurvature perturbation modes 
is illustrated in Fig.~\ref{fig:compare}, where the red line shows the unique adiabatic slope and the blue shading indicates the range of possible isocurvature slopes as $\alpha$ varies from 0 to 1. The clear separation between these regions establishes the spectral slope as a robust discriminant between isocurvature and adiabatic perturbations, providing a powerful observational signature for probing primordial fluctuation modes.

\section{Conclusion and discussion}\label{conclusion}

We have analyzed the spectral characteristics of IGWs from primordial perturbations, focusing on discriminating features between adiabatic and isocurvature modes. Our analysis demonstrates that these modes, governed by distinct equations of motion, produce characteristic signatures in the SGWB. For isocurvature IGWs, we find that the infrared spectral slope takes the form shown in Eq.~(\ref{eq:nGWb}), 
while for narrow spectra near the peak, a more refined expression (Eq.~\ref{eq:nGWimprove})) is obtained. 
In contrast, the adiabatic mode produces IGWs with different spectral slope~(Eq.~(\ref{nGW35})). 

From an observational perspective, it is important to note that the peak scale of $\Omega_{\text{GW}}(k)$ generally differs from that of $\mathcal{P}(k)$ due to the explicit $k$ dependence in Eq.~\eqref{Ptilde}. This peak shifts toward smaller scales, approximately coinciding with the peak of $\widetilde{\mathcal{P}}(k)$, as we have demonstrated for both lognormal and broken power spectra through Eqs.~\eqref{eq:ke1} and \eqref{eq:ke2}.

Our results establish distinctive spectral signatures for IGWs from isocurvature perturbations, providing a potential observational discriminant between adiabatic and isocurvature primordial modes through their GW imprints.
 
\begin{acknowledgments}
We thank Lang Liu and Chen Yuan for useful discussions that greatly improved our manuscript.
This work was supported by the National Key Research and Development Program of China (Grant No.~2020YFC2201502),
the National Natural Science Foundation of China (Grants No.~12405056, No.~12275080, No.~12203004, and No.~12075084), and the Innovative Research Group of Hunan Province (Grant No.~2024JJ1006).
\end{acknowledgments}

\bibliography{ref}	
\end{document}